\documentclass[a4paper,12pt]{article}
\usepackage[latin1]{inputenc}
\usepackage{float}
\usepackage{amsmath,amssymb}
\usepackage{theorem}
\usepackage{graphicx}
\usepackage{color}
{\theorembodyfont{\sffamily} \newtheorem{theorem}{Theorem}
{\theorembodyfont{\slshape} \newtheorem{definicion}{Definition}

\begin{document}
\title{\bf Characterizing exact solutions from asymptotic physical concepts}  
\author{
Alejandro Perez${}^{1,}$\thanks{present address: 
Centre de Physique Th\'eorique, CNRS Luminy, 
             F-13288 Marseille, France.
email: aperez@phyast.pitt.edu}\;
and Osvaldo M. Moreschi${}^{2,}$\thanks{Member of CONICET. 
email: moreschi@fis.uncor.edu} \\
${}^1$\small Department of Physics and Astronomy, University of Pittsburgh, \\
\small Pittsburgh, PA 15260, USA\\   
${}^2$\small FaMAF, Universidad Nacional de C\'ordoba, \\
\small Ciudad Universitaria, 5000 C\'ordoba, Argentina\\
} 
\date{December 27, 2000}

\maketitle

\begin{abstract}
We contribute to the subject of the physical interpretation of exact
solutions by characterizing them through a systematic study in terms of
unambiguous physical concepts coming from systems in linearized gravity.
We use the physical meaning of the leading order behavior of the Weyl
spinor components $\Psi_0^0$, $\Psi_1^0$ and $\Psi_2^0$ and 
of the Maxwell spinor components $\phi_0^0$ and $\phi_1^0$
and integrate from future null infinity inwards the 
exact field equations. In this way it is assigned an unambiguous
physical meaning to exact solutions and we indicate a method to 
generalize the procedure to radiating spacetimes.
\end{abstract}


\section{Introduction}

The problem of constructing an adequate dynamical description of a system of
gravitating compact objects has received much attention recently. Several
different approaches can be distinguished to tackle this problem. 
They range from post-Newtonian schemes, which can handle systems with low
relative velocities, to numerical methods that confront the problem of
solving the full Einstein equations;
by evolving some given initial data on a
spacelike hypersurface. 

There are however other approaches that intend to use mainly an evolution in
terms of characteristic surfaces. In the past we have suggested a set of
sections at future null infinity($\cal I ^{+}$), the so called 
``nice sections''\cite{Moreschi88}, 
that define asymptotically, in the interior of the spacetime, a
set of appropriate characteristic surfaces for this task. Very recently we
have proved that this sections exist globally on $\cal I^{+}$, and that they
have the expected correct behavior\cite{Moreschi98}\cite{Dain99'}. 

%
In all these approaches it is important to know how to characterize the
physical system one has in mind in terms of the physical fields.
For example, there is growing interest in the study of collisions of compact
objects and the generation of gravitational radiation in these processes.
If one tries to describe its evolution in terms of characteristic data,
one is faced with the problem of ascribing to the asymptotic physical
fields the information of the physical system in the interior of 
the spacetime.

In this article we would like to contribute on the understanding of our
approach by concentrating on the problem of the physical meaning of the
asymptotic fields at future null infinity. The point is that if one has a
preferred set of sections at $\cal I^{+}$, or equivalently a preferred set of
characteristic hypersurfaces, one should also know how to read or introduce
initial data with the desired physical meaning. We know how to ascribe
unambiguous physical meaning to the fields in the context of linearized
gravity; however one encounters some difficulties
when one wants, for example, to give an unambiguous physical meaning to
total angular momentum in a spacetime admitting gravitational 
radiation\cite{Moreschi86}\cite{Moreschi2000}.

One possibility for the study of a system with compact objects is to try to
build a framework that uses a finite set of degrees of freedom; which are
determined by the finite parameters determining the compact objects. In this
approach one would need to know how to read this structure from the physical
fields, in particular, the asymptotic geometric fields at future null
infinity. More concretely, if the system consists of two compact objects;
which are characterized by their respective mass, momentum, intrinsic
angular momentum and their relative position; one would need to relate this
structure, of the interior of the spacetime, with the structure of the
asymptotic fields. The present work is a contribution on this relation.

In some realistic astrophysical systems there is an initial era in which the
gravitational radiation is very small; as for example is the case of a
coalescent binary system. When the compact objects are
very far apart and with small relative velocities, one would like to resort
to properties of each compact object. In this regime, the quantities 
describing these properties, should have a clear counterpart in the
corresponding description of the system in linearized gravity. And it should
be noted that the concepts constructed in linearized gravity have a natural
extension to non-radiating spacetimes; where the supertranslation gauge
freedom can be fixed. The first task is then to understand the relation
between the structure of the sources and that of the asymptotic fields when
there is no gravitational radiation.

Consequently, in this paper we start to study the issue by restricting
ourselves to the case of stationary spacetimes.
We  construct a relation between the structure of the asymptotic fields,
for systems with compact objects, with the structure of the internal fields
coming from Einstein equations. More concretely, we integrate ``inwards'',
from $\cal I^{+}$, the exact field equations from appropriate asymptotic data.
These asymptotic data are suggested by the analysis of the
corresponding system in linearized gravity, where one has the unambiguous
physical interpretation provided by the background metric.

We  make extensive use of the null tetrad formalism\cite{NPF}, following
the notation defined by Geroch, Held, and Penrose in \cite{Geroch73} (which we
refer to as GHP). The explicit form of the spin coefficients, Einstein
equations, and the Bianchi identities can be found in \cite{Geroch73} and
\cite{Moreschi87}. In the next section we define the null tetrads upon which
our calculations are based. Two definitions of tetrads and associated
coordinate systems are given for asymptotically flat spacetimes.

In section \ref{lili} we analyze the properties of asymptotic fields in
linearized gravity. In subsections \ref{sweil} and \ref{slinemax} we study the
asymptotic structure of the Weyl tensor, and the Maxwell fields respectively.

Motivated by the results of \ref{sweil} we give, in section \ref{sexact},
the definitions of {\em monopolar spacetimes} and 
{\em monopolar spacetimes with angular momentum}. 
We find all the exact solutions satisfying these definitions.

In section \ref{sexactmax} we generalize the results of the previous section 
to the case
of Einstein-Maxwell solutions, and we give the definitions of {\em monopolar
spacetimes with charge} and {\em monopolar spacetimes with angular momentum
and charge} respectively. We also find the exact solutions satisfying these
last definitions.

\section{\label{tetrads}Standard tetrads and rotating tetrads of future
asymptotically flat space-times}

\subsection{Standard Tetrad\label{standard}}

Our studies are all concerned with asymptotically flat spacetimes at future
null infinity. The first
task is to define the appropriate null tetrads in terms of which we  express
Einstein equations in the GHP formalism. It is convenient to introduce the
notion of {\em standard} and {\em rotating null tetrads} which are defined in terms of
the asymptotic structure of the spacetime.

In Appendix \ref{sappenda}, we present the basic equations
associated to a general null tetrad adapted to an arbitrary null congruence,
where it is shown how a local coordinate system can be constructed based on
an appropriate hypersurface $\Sigma $. In this section we make the analogous
construction taking null infinity as our preferred hypersurface. We assume
null infinity (${\cal I}^{+}$) to have the topology of 
$S^{2}\times \mathbb R $\cite{Moreschi87}. 
Therefore, we can make use of a coordinate system $\left( 
\tilde{u},\tilde{x}^{2},\tilde{x}^{3}\right) $ on ${\cal I}^{+}$ (where $%
\tilde{u}$ corresponds to a coordinate along the generators of $\cal I^{+}$
and $\tilde{x}^{2}\,,\,\tilde{x}^{3}$ coordinatize the sphere $S^{2}$). In
this way the equation $\tilde{u}={\tt constant}$ determines a family of smooth
sections on $\cal I^{+}$ which we shall denote by $S_{u}$. Conversely given
any smooth family of sections on $\cal I^{+}$, labeled by $\tilde{u}$, one
can construct the coordinate system $\left( \tilde{u},\tilde{x}^{2},\tilde{x}%
^{3}\right) $ at $\cal I^{+}$, such that the coordinates 
$\left( \tilde{x}^{2},\tilde{x}^{3}\right) $ are constant along the 
generators of $\cal I^{+}$.

Let $S_{u}$ be a given smooth family of sections of $\cal I^{+}$. One can
extend the coordinate system $\left( \tilde{u},\tilde{x}_{2},\tilde{x}%
_{3}\right) $ into the interior of the spacetime by considering the null
geodesics in the interior reaching each $S_{u}$ orthogonally. For each $%
\widetilde{u}$, this family of null geodesics generates a null surface $N_{u}$%
; we define the function $u$ in the interior of the spacetime such that it
is constant on each $N_{u},$ and $u=\tilde{u}$ at $\cal I^{+}$. Similarly, we
define the coordinates $\left( x^{2},x^{3}\right) $ in a neighborhood of 
$\cal I^{+}$, such that they are constant along the generators of $N_{u}$, 
and $x^{2}=\tilde{x}^{2}$ and $x^{3}=\tilde{x}^{3}$ at $\cal I^{+}$. The fourth
coordinate $r$ is taken as an affine parameter along the generators of 
$N_{u} $. In this manner we obtain a suitable coordinate system $\left(
u,r,x^{2},x^{3}\right) $ in a neighborhood of $\cal I^{+}$. By construction
this null congruence is surface-forming, i.e., the null vector field $\ell
_{a}=\left( du\right) _{a}$ is tangent to the null geodesics for which $r$
is an affine parameter, i.e., 
$\ell ^{a}=\left( \frac{\partial }{\partial \,r}\right) ^{a}$. 
Latin indices will be understood as abstract indices unless
otherwise stated.

Now we construct the null tetrad $(\ell _{a},n_{a},m_{a},\bar{m}_{a})$
following the path described in appendix \ref{sappenda}.

In order to complete the definition of the null tetrad, that we shall call 
{\em standard}, additional conditions must be required. First we impose that
the vector $m^a$ (and its complex conjugate) satisfies
\begin{equation}
m\left( u\right) =m\left( r\right) =0.  \label{eeme}
\end{equation}
This is always possible by means of null rotations of type I and II, see
appendix \ref{gauge}.

Then we require the coordinate system $\left( \tilde{u},\tilde{x}^{2},\tilde{%
x}^{3}\right) $ at infinity to be Bondi-like; i.e., such that the
restriction of the conformal metric tensor to $\cal I^{+}$ is given by
\begin{equation}
\tilde{g}^{+}=-\frac{4d\zeta d\bar{\zeta}}{\left( 1+\zeta \bar{\zeta}\right)
^{2}},
\end{equation}
where we are using the stereographic coordinates $\zeta =\frac{1}{2}\left( 
\tilde{x}^{2}+i\tilde{x}^{3}\right)$ on the sphere.

Under a tetrad rotation of the form $m^{a}\rightarrow e^{i\theta }m^{a}$ the
spin coefficient $\epsilon $ transforms in the following way:
\begin{equation}
\epsilon \rightarrow \epsilon +\frac{1}{2}i\ell \left( \theta \right) .
\end{equation}
The spin coefficient $\epsilon $ is pure imaginary, since $\ell ^{a}$ is
geodesic\footnote{%
The real part of $\epsilon $ is proportional to $n^{a}\ell ^{b}\nabla
_{b}l_{a}$.(see definition of the spin coefficients in \cite{Geroch73})};
therefore, it is possible to choose $\theta $ such that $\epsilon $
vanishes. The second condition is then $\epsilon =0$. There is still a
freedom in the choice of $\theta $, namely, we can add any arbitrary
function $\theta _{0}$ independent of $r$. This freedom represents a
rotation at infinity which allows us to take
\begin{equation}\label{eq:masymp}
m^{a}=\frac{\sqrt{2}P_{0}}{r}\left( \frac{\partial }{\partial \zeta }\right)
^{a}+O\left( \frac{1}{r^{2}}\right) ,
\end{equation}
where $P_{0}=\frac{1}{2}\left( 1+\zeta \bar{\zeta}\right) $.

Finally, the affine parameter $r$ is undetermined by an additive function $%
r_{0}\left( u,x^{2},x^{3}\right)$; which we fix by the requirement
\begin{equation}
\rho =-\frac{1}{r}+O\left( \frac{1}{r^{3}}\right) ;  \label{roro}
\end{equation}
which is equivalent to think of $r$ as the luminosity distance. This
choice of $r$ not only simplifies $\rho $, but also gives it physical
relevance.

The leading terms $X_{0}^{2}$ and $X_{0}^{3}$ (which are independent of $r$), 
of the asymptotic expansion
of the components of the null vector $n^{a}$ (see  (\ref{vecn})), 
vanish because of the
definition of the asymptotic coordinates and the fact that the vector field $%
n^{a}$ is tangent to $\cal I ^{+}$. It can be seen that $X^{0}_0$ has the unit
value in this case; due to the fact that $n^{a}\ell _{a}=1$ and $\ell =du$%
\cite{Moreschi87}.

In appendix \ref{sappenda} we present the equations imposed by the
torsion-free and metric conditions on the connection for a general tetrad
associated to a null congruence. The corresponding relations for a {\em %
standard tetrad} are easily obtained by setting $\omega =0$, $\xi ^{0}=0$, $%
\rho =\bar{\rho}$, $\kappa =0$, and $\epsilon =0$ in those equations.

\subsection{Rotating Tetrad\label{rotating}}

Although it is always possible to construct a {\em standard tetrad} in a
neighborhood of null infinity for any asymptotically flat spacetime, in
certain cases it will be more convenient to work with null tetrads adapted
to a twisting congruence of null geodesics. In this section we define the
notion of {\em rotating tetrad}, as a particularization of the general tetrad
given in appendix \ref{sappenda}.

The fundamental condition imposed on the {\em rotating tetrad} is the
vanishing of the leading asymptotic term of one of the Weyl tensor components,
namely  
\begin{equation}
\lim_{r\rightarrow \infty }\,\,\,r^{4}\psi _{1}\equiv \psi
_{1}^{0}=0,
\end{equation} 
where $\psi _{1}=C_{abcd}\ell^{a}n^{b}\ell^{c}m^{d}$. This can
be achieved by noting that at the section $S_{u}$ one can construct a
congruence of null geodesics such that $\psi _{1}^{}=0$ on $S_{u}$ by
appropriately choosing the corresponding twist of the congruence at $\cal
I^{+}$. The resulting congruence fails to be surface forming ($\rho
-\bar{\rho}\neq 0$). Additional conditions must be imposed in order to
completely determine the tetrad and it associated coordinate system.

As in the case of the {\em standard tetrad} we take Bondi-coordinates $%
\left( \tilde{u},\tilde{x}_{2},\tilde{x}_{3}\right) $ at $\cal I^{+}$. The
'origin' of the coordinate $r$ (affine parameter along the twisting
geodesics) it is chosen in such a way that
\begin{equation}
\rho +\bar{\rho}=-\frac{2}{r}+O\left( \frac{1}{r^{3}}\right) .  \label{ro}
\end{equation}
This condition is analogous to the one imposed to the {\em standard tetrad}
in eq. (\ref{roro}). 

It is again possible to choose the vector $m^{a}$ 
satisfying $m\left( r\right) =0$ (see appendix \ref{gauge}). In this 
case the tetrad can
also be chosen such that $\epsilon =0$, in an analogous way as in the previous
case for the {\em standard tetrad}. The remaining freedom encoded in $\theta
_{0}$, in the transformation $m^{a}\rightarrow e^{i\theta }m^{a}$, is fixed
by requiring
\begin{equation}
\xi _{0}^{2}=\frac{\sqrt{2}P_{0}}{r}+O\left( \frac{1}{r^{2}}\right),
\label{exi0}
\end{equation}
\begin{equation}
\xi _{0}^{3}=-i\frac{\sqrt{2}P_{0}}{r}+O\left( \frac{1}{r^{2}}\right) ;
\label{exi0p}
\end{equation}
which is equivalent to condition (\ref{eq:masymp}).
From the definition of the asymptotic coordinate system, one can show that the
leading terms in the asymptotic expansion of the components of $n^a$ take the
values \begin{equation} X_{0}^{0}=1,\;\;\;X_{0}^{2}=0,\;\;\;X_{0}^{3}=0, 
\label{x} \end{equation}
for any asymptotically flat spacetime. The torsion-free and metric
conditions for the {\em rotating tetrad} can be obtained, from the general
equations appearing in the appendix \ref{sappenda} by setting $\omega =0$, 
$\kappa =0$, and $\epsilon =0$. Finally let us point out that the notion of 
{\em rotating tetrad }exists for any asymptotically flat spacetime, and
coincides with that of the {\em standard tetrad} when $\psi _{1}^{0}$
vanishes.

\section{Linearized gravity\label{lili}}

\subsection{The Weyl Tensor in Stationary Asymptotically Flat Space-times
(Vacuum case)\label{sweil}.}

In this section we analyze the asymptotic structure of the Weyl tensor in
linearized gravity for isolated systems. The results obtained here will
provide physical motivation for the asymptotic conditions required by the
definitions that will be made in sections \ref{sexact} and \ref{sexactmax}. 
The study is based
on the restrictions imposed on the curvature by the Bianchi identities.
These restrictions define the multipolar structure of the geometry of an
isolated system in linearized gravity.

We  use a null coordinate system, and the corresponding null tetrad 
$( \ell ^{a}, n^{a},\\ m^{a}, \bar{m}^{a})$ adapted to a congruence of
hypersurface-orthogonal, shear free, null geodesics of Minkowski spacetime.
These conditions imply that this null coordinate system is the one defined
by the null cones emanating from an arbitrary world line $\gamma (u)$ given
by the equation $x^{\mu }=z^{\mu }\left( u\right) $; where $\mu =0,1,2,3$.
In our case we take $z^{\mu }\left( u\right) =u\delta _{0}^{\mu }$. The
usual Cartesian coordinates $(x^{\mu })=\left( t,x,y,z\right) $ are related to
the null polar coordinates $\left( u,r,\zeta ,\bar{\zeta}\right) $ by $%
x^{\mu }=z^{\mu }\left( u\right) +r\,l^{\mu }\left( \zeta ,\bar{\zeta}%
\right) $, where $l^{\mu }\left( \zeta ,\bar{\zeta}\right) $ is the null
vector in Minkowski spacetime pointing in the direction given by $\zeta $
and $\bar{\zeta}$, where $(l^{\mu })=\frac{1}{2P_{0}}( 1+\bar{\zeta}\zeta
,\zeta +\bar{\zeta},i( \bar{\zeta}-\zeta ) , -1+\zeta \bar{\zeta}%
) $. In terms of the above coordinate system the line element is $%
ds^{2}=du^{2}+2dudr-r^{2}\frac{d\zeta d\bar{\zeta}}{P_{0}^{2}}$, and the
null tetrad becomes: 
\begin{equation}
\ell ^{a}=\left( \frac{\partial }{\partial r}\right) ^{a},
\end{equation}
\begin{equation}
n^{a}=\left( \frac{\partial }{\partial u}\right) ^{a}-\frac{1}{2}\left( 
\frac{\partial }{\partial r}\right) ^{a},
\end{equation}
\begin{equation}
m^{a}=\frac{\sqrt{2}P_{0}}{r}\left( \frac{\partial }{\partial \zeta }\right)
^{a}.
\end{equation}

The non-vanishing spin coefficients corresponding to our null tetrad are: $%
\rho _{M}=-\frac{1}{r}$, $\rho _{M}^{\prime }=\frac{1}{2r}$, and $\beta _{M}=%
\bar{\beta}_{M}^{\prime }=-\frac{1}{r\sqrt{2}}\frac{\partial P}{\partial
\zeta }$.

We assume that the energy-momentum tensor of linearized gravity is
stationary and has compact support; therefore, the field equations have no
sources in a neighborhood of $\cal I^{+}$.

The linearized Bianchi identities depend only on the spin coefficients of
Minkowski spacetime. In a regular asymptotically flat spacetime\cite{Moreschi87}
the Weyl components $\psi _{j}^{{}}$ with $(j=0,1,2,3,4)$ are given by the
following asymptotic series in negative powers of $r$
\begin{equation}
\psi _{j}=\sum\limits_{n=0}^{{}}\frac{\psi _{j}^{n}}{r^{\left( n+5-j\right) }},
\end{equation}
where the $\psi _{j}^{n}$ are independent of $r$.

The radial Bianchi identities, (see equations (3.88-3.91) in \cite
{Moreschi87}), restrict the asymptotic series in $r$ for the
components of the Weyl tensor to: \begin{equation}
\psi _{1}=\frac{\psi _{1}^{0}}{r^{4}}-\frac{\overline{\eth }_{0}\psi _{0}^{0}%
}{r^{5}}-\frac{1}{2}\frac{\overline{\eth }_{0}\psi _{0}^{1}}{r^{6}}-\frac{1}{3%
}\frac{\overline{\eth }_{0}\psi _{0}^{2}}{r^{7}}-\frac{1}{4}\frac{\overline{%
\eth }_{0}\psi _{0}^{3}}{r^{8}}+\ldots \label{eps1},
\end{equation}
\begin{equation}
\psi _{2}=\frac{\psi _{2}^{0}}{r^{3}}-\frac{\overline{\eth }_{0}\psi _{1}^{0}%
}{r^{4}}+\frac{1}{2}\frac{\overline{\eth }_{0}^{2}\psi _{0}^{0}}{r^{5}}+\frac{%
1}{2\times 3}\frac{\overline{\eth }_{0}^{2}\psi _{0}^{1}}{r^{6}}+\frac{1}{%
3\times 4}\frac{\overline{\eth }_{0}^{2}\psi _{0}^{2}}{r^{7}}+\ldots
\label{eps2},
\end{equation}
\begin{equation}
\psi _{3}=\frac{\psi _{3}^{0}}{r^{2}}-\frac{\overline{\eth }_{0}\psi _{2}^{0}%
}{r^{3}}+\frac{1}{2}\frac{\overline{\eth }_{0}^{2}\psi _{1}^{0}}{r^{4}}-\frac{%
1}{2\times 3}\frac{\overline{\eth }_{0}^{3}\psi _{0}^{0}}{r^{5}}+\frac{1}{%
2\times 3\times 4}\frac{\overline{\eth }_{0}^{3}\psi _{0}^{1}}{r^{6}}+\ldots
\label{eps3},
\end{equation}
\begin{equation}
\psi _{4}=\frac{\psi _{4}^{0}}{r}-\frac{\overline{\eth }_{0}\psi _{3}^{0}}{%
r^{2}}+\frac{1}{2}\frac{\overline{\eth }_{0}^{2}\psi _{2}^{0}}{r^{3}}-\frac{1%
}{2\times 3}\frac{\overline{\eth }_{0}^{3}\psi _{1}^{0}}{r^{4}}+\frac{1}{%
2\times 3\times 4}\frac{\overline{\eth }_{0}^{4}\psi _{0}^{0}}{r^{5}}+\ldots
\label{eps4},
\end{equation}
where the operators $\eth _{0}$,
and $\overline{\eth }_{0}$ represent the edth operators on the unit 
sphere\cite{Newman66}\cite{Goldberg67} in the GHP notation\cite{Geroch73}.

New restrictions on the Weyl components come from the non-radial Bianchi
identities, corresponding to the primed eqs.(3.88-3.93) in \cite{Moreschi87}.
These restrictions fix the angular dependence of the Weyl components in the
following way:
\begin{equation}
\psi _{0}=\sum\limits_{\ell =2}^{{}}\sum\limits_{m=-\ell }^{\ell }\frac{%
a^{\ell m}}{r^{\ell +3}}\,_{2}Y_{lm},
\end{equation}
\begin{equation}
\psi _{1}^{0}=\sum\limits_{m=-1}^{1}b^{m}\,_{1}Y_{1m}^{{}},
\end{equation}
\begin{equation}
\psi _{2}^{0}=-M,
\end{equation}
where the $_{s}Y_{\ell \,m}$ are generalized spherical harmonics of spin
weight $s$\cite{Newman66}, and $a^{\ell m},\,b^{m}$ together with $M$ are
constants. From the stationarity requirement and the Bianchi identities one
can see that the remaining leading order terms $\psi _{3}^{0}$ and $\psi
_{4}^{0}$ vanish\cite{Moreschi87}.

This constitutes the multipolar structure of stationary isolated
systems in linearized gravity.

\subsection{The Electromagnetic Tensor\label{slinemax}}

In linearized Einstein-Maxwell theory Maxwell
equations are directly solved on the flat Minkowski background. In this
section we recall the well known result of the multipolar decomposition of
the electromagnetic field in Minkowski spacetime in terms of the null tetrad
formalism. The results will be used in section \ref{sexactmax} 
to define the asymptotic
data of the Maxwell fields for the exact Einstein-Maxwell equations.

The components of the electromagnetic tensor in the null tetrad are $\phi
_{0}\equiv F_{ab}\ell ^{a}m^{b}$, $\phi _{1}\equiv \frac{1}{2}F_{ab}\left(
\ell ^{a}n^{b}+\bar{m}^{a}m^{b}\right) $ and $\phi _{2}\equiv F_{ab}\bar{m}%
^{a}n^{b}$. The radial dependence of the electromagnetic field of compact
sources is determined by the un-primed Maxwell equations in the
Newman-Penrose formalism\cite{Geroch73}\cite{Newman80}. 
If we denote the asymptotic series of $%
\phi _{0}$ by 
\begin{equation}
\phi _{0}=\sum \frac{\phi _{0}^{\,i}}{r^{3+i}}, 
\end{equation}
then the
other components are given by:
\begin{equation}
\phi _{1}=\frac{q}{r^{2}}-\frac{\overline{\eth}_0\phi _{0}^{0}}{r^{3}}-\frac{1%
}{2}\frac{\overline{\eth}_0\phi _{0}^{1}}{r^{4}}-\frac{1}{3}\frac{\overline{%
\eth}_0\phi _{0}^{2}}{r^{5}}\cdots,
\end{equation}
\begin{equation}
\phi _{2}=\frac{\phi _{2}^{0}}{r}-\frac{\overline{\eth}_0\phi _{1}^{0}}{r^{2}}+%
\frac{1}{2}\frac{\overline{\eth}_0^{2}\phi _{0}^{1}}{r^{3}}+\frac{1}{6}\frac{%
\overline{\eth}_0^{2}\phi _{0}^{2}}{r^{4}}\cdots
\end{equation}

Finally, the solution of all the Maxwell equations for the case of stationary
compact sources has the following structure:
\begin{equation}
\phi _{0}=\frac{\sum \mu ^{m}\,_{1}Y_{1m}}{r^{3}}+\frac{\sum
Q^{m}\,_{1}Y_{2m}}{r^{3}}+\ldots,
\end{equation}
\begin{equation}
\phi _{1}=\frac{q}{r^{2}}%
+\ldots,
\end{equation}
\begin{equation}
\quad\phi _{2}=O\left( \frac{1}{r^{3}}\right),
\end{equation}
where $\mu^{m} $, $Q^{m}$, and $q$ are constants, corresponding 
to the dipolar moment,
quadrupolar moment, and charge of the Maxwell field respectively. 
The component $\phi_{2}^{0}$ corresponds to electromagnetic radiation. 
In the stationary case one has $\phi _{2}^{0}=0$.

\section{Vacuum exact solutions\label{sexact}}

In section \ref{sweil} we studied the asymptotic structure of the Weyl
tensor required by the Bianchi identities in linearized gravity. The
structure of a compact object in linearized gravity appears in the
asymptotic expansion of the Weyl tensor as a multipolar series in inverse
powers of the radial coordinate. Now we look for exact solutions of
Einstein equations that have the analogous asymptotic structure. The
vacuum exact solutions that are characterized by the restrictions imposed by
the asymptotic structure of compact sources in linearized gravity are given
by the following definitions.

\subsection{Monopolar spacetime}
\begin{definicion}\label{defmono}
 We call a spacetime {\em monopolar} if it is a
stationary, vacuum asymptotically flat, solution of Einstein equations
and, in addition, the following conditions hold in the {\em standard tetrad}
defined in section (\ref{standard}), namely: $\psi _{0}=\psi _{1}^{0}=0$,
and $\psi _{2}^{0}\neq 0$.
\end{definicion}
The conditions $\psi _{0}=0$ and $\psi _{1}^{0}=0$ correspond to the
requirement that the spacetime does not have further structure than mass
(recall that according to our results of section (\ref{sweil}) in linearized
gravity $\psi _{0}^{{}}$ is where quadrupolar and higher momenta are
encoded, while $\psi _{1}^{0}$ corresponds to the angular momentum data).

It turns out that the definition of monopolar spacetimes is strong enough to
single out a family of solutions of the Einstein equations. The result is
expressed in the following theorem.
\begin{theorem}\label{teomono}
 The solutions to the vacuum Einstein equations which are
stationary, asymptotically flat, and in the {\em standard tetrad} $\psi
_{0}=\psi _{1}^{0}=0$, and $\psi _{2}^{0}\neq 0$, are given by the
one-parameter family of Schwarzschild spacetimes. Alternatively, a spacetime
is {\em monopolar}, in the sense given by the definition above, if and only
if it belongs to the one-parameter family of Schwarzschild geometries.
\end{theorem}
{\bf Proof:} We start with Sachs equations, the optical scalars equations
in the GHP formalism, that in this case reduce to:
\begin{equation}
\frac{\partial \rho }{\partial r}=\rho ^{2}+\sigma \bar{\sigma}
\quad,\quad\frac{%
\partial \sigma }{\partial r}=2\rho \sigma.
\end{equation}

This set of equations are solved by
\begin{equation}
\sigma =\frac{\sigma _{0}}{r^{2}-\sigma _{0}\bar{\sigma}_{0}}
\quad,\quad\rho =\frac{-r}{r^{2}-\sigma _{0}\bar{\sigma}_{0}};
\end{equation}
where $\sigma_0$ does not depend on $r$.

The Bianchi  identity involving the radial derivative of $\psi _{1}$\cite
{Moreschi87} reduces to
\begin{equation}
\frac{\partial \psi _{1}}{\partial r}=4\rho \psi _{1};
\end{equation}
which implies that $\psi _{1}=\frac{\psi _{1}^{0}}{(r^{2}-\sigma _{0}\bar{%
\sigma}_{0})^{2}}$. The definition of monopolar spacetime requires that $%
\psi _{1}^{0}=0$; therefore, $\psi _{1}$ must vanish. At this point, the
hypothesis of the Goldberg-Sachs theorem are fulfilled, and consequently the
shear $\sigma $ must vanish. This implies that the spacetime is shear free
algebraically special of type II. The only vacuum shear free, algebraically
special solution of type II are the Robinson-Trautman spacetimes\cite{Newman62};
and among the Robinson-Trautman solutions the only stationary ones are the
Schwarzschild geometries.$\square $

\subsection{Monopolar spacetime with angular momentum}

The next step is to introduce angular momentum in our definitions. The
existence of angular momentum is given, in the {\em standard tetrad}, by a
non-vanishing $\psi _{1}^{0}$. As we pointed out in our definitions of null
tetrads in this case we have the possibility of defining a spacetime with
angular momentum in either the standard tetrad or the rotating tetrad, since
they differ when $\psi _{1}^{0}\neq 0$. In this section we  study the
implications of both alternatives in the following definitions of monopolar
spacetimes with angular momentum.
\begin{definicion}
 A spacetime is called {\em monopolar with angular
momentum (type a)} when it is a stationary, vacuum, locally 
asymptotically flat%
\footnote{%
By locally asymptotically flat it is meant that the conditions for
asymptotic flatness of \cite{Moreschi87} are satisfied locally at 
$\cal I^{+}$.
In particular it might be that future null infinity is not complete.}
solution of Einstein equations, and in addition, its Weyl tensor in the 
{\em standard tetrad} satisfies that\thinspace $\psi _{1}^{0}\neq 0,\,\,\psi
_{2}^{0}\neq 0,\,\,\psi _{0}=0$.
\end{definicion}

Starting with the asymptotic conditions imposed by the previous definition
one proceeds to integrate inwards the Einstein equations from future null
infinity. The result of these calculations are presented in the following
theorem.
\begin{theorem}
The spacetimes which are {\em monopolar with angular
momentum (type a)} are the stationary Newman-Tamburino solutions.
\end{theorem}
{\bf Proof:} One can easily see that the definition 2a implies
\begin{equation}
\psi _{0}=\rho -\bar{\rho}=0,
\end{equation}
\begin{equation}
\rho ^{2}\neq \sigma \bar{\sigma}\quad,\quad\rho \neq 0,
\end{equation}
\begin{equation}
  \label{eq:rotl}
  \nabla_{[a}\ell_{b]}=0,
\end{equation}
\begin{equation}
\psi _{1}^{0}\neq 0.
\end{equation}

These equations constitute the complete characterization of the
Newman-Tamburino solutions\cite{Newman62}. Since the definition requires the
spacetime to be stationary, one concludes that the spacetimes which are {\em %
monopolar with angular momentum of type a} are determined by the stationary
Newman-Tamburino solutions.

These solutions are not asymptotically flat in all null future directions 
and therefore do not describe
the type of physical systems in which we are interested. This result
suggests that the twist-free {\em standard tetrad} might not be well suited
to the study of spacetime geometries with angular momentum. In the following
definition we apply our strategy making use of the {\em rotating tetrad}.
\begin{definicion}
 A spacetime is called {\em monopolar with angular
momentum (type b)} when it is a stationary, vacuum, locally asymptotically
flat solution of Einstein equations, and, in addition, its Weyl tensor in
the {\em rotating tetrad} satisfies that\thinspace \thinspace 
$\psi_{2}^{0}\,\neq 0$ and $\psi _{0}=0$\footnote{In both definitions we take 
$\psi _{0}=0$ as we did in Definition \ref{defmono} in
order to introduce no further multipole moments.}.
\end{definicion}
These spacetimes are characterized by:
\begin{theorem}\label{tangb}
The spacetimes which are {\em monopolar with angular
momentum (type b)} correspond to the spherical Kerr family. The following
are the expressions for the components of the null tetrad.
\begin{equation}
\xi ^{i}=\frac{\xi _{0}^{i}}{\left( r-ia\cos \left( \theta \right) \right) },
\end{equation}
\begin{equation}
X^{i}=\delta _{0}^{i}-\frac{\left( \tau _{0}\bar{\xi}_{0}^{i}+\bar{\tau}%
_{0}\xi _{0}^{i}\right) }{\left( r^{2}+a^{2}\cos ^{2}\left( \theta \right)
\right) },
\end{equation}
\begin{equation}
U=-\frac{1}{2}-\frac{\tau _{0}\bar{\tau}_{0}}{\left( r^{2}+a^{2}\cos
^{2}\left( \theta \right) \right) }+\frac{Mr}{\left( r^{2}+a^{2}\cos
^{2}\left( \theta \right) \right) },
\end{equation}
with $\xi _{0}^{0}=-\tau _{0}=i\eth _{0}\left( a\cos \left( \theta \right)
\right) $. The parameters $M$ and $a$ are the mass and angular
momentum parameter respectively.
\end{theorem}
We refer the reader to the Appendix \ref{sappendb} where this theorem is
proved.

\section{Einstein-Maxwell exact solutions\label{sexactmax}}

Our construction for vacuum spacetimes can be generalized to the electrovac
case using the same idea of giving asymptotic data defined by
the study of fields in the linear theory where their physical interpretation is
available. In section \ref{slinemax} we have described the structure of the
asymptotic electromagnetic field in  terms of
a null tetrad. Previously we have showed, in the vacuum case, that imposing
analogous asymptotic conditions we can single out certain families of exact
solutions. In this section we  continue with the program, and combining
the linear analysis of Weyl and electromagnetic tensors we give new
definitions of Einstein-Maxwell space-times which generalizes the previous
results.

In the following definition we start by adding charge to the  Monopolar
spacetimes. \begin{definicion}\label{defmonoq}
We define a spacetime to be {\em monopolar with charge}
when it is a stationary asymptotically flat solution of Einstein-Maxwell
equations such that in the {\em standard tetrad} its Weyl and
electromagnetic tensor satisfy the following asymptotic conditions: $\psi
_{0}=\psi _{1}^{0}=0$, $\psi _{2}^{0}\neq 0$ and $\phi _{0}=0,\,\,\,\phi
_{1}^{0}\neq 0$.
\end{definicion}
The condition $\phi _{0}=0$ rules out higher multipole moments in the
electromagnetic field in the spirit of the linearized analysis.

The restrictions imposed by the previous definitions are reflected by the
following theorem. \begin{theorem}\label{teomonoq}
The stationary, asymptotically flat solutions of the
Einstein-Maxwell equations such that in the {\em standard tetrad} $\psi
_{0}=\psi _{1}^{0}=0$ , $\psi _{2}^{0}\neq 0$ , $\phi _{0}=0,$ and $\phi
_{1}^{0}\neq 0$ are given by the two-parameter family of Reissner-Nordstrom
spacetimes.
Alternatively, using the definition given above, a spacetime is {\em %
monopolar with charge} if and only if it belongs to the Reissner-Nordstrom
family.
\end{theorem}

We will give a full proof of the following theorem which generalizes the
previous one. In the sequel we introduce charge to the Monopolar spacetimes
with angular momentum. 

\begin{definicion}\label{defmonoqang}
We define a spacetime to be monopolar with charge and
angular momentum when it is a stationary asymptotically flat solution of
Einstein-Maxwell equations such that in the {\em rotating tetrad} its Weyl
and electromagnetic tensor satisfy the following asymptotic conditions: $%
\psi _{0}=0$, $\phi _{0}=0$, and $\phi _{1}\neq 0$.
\end{definicion}

These spacetimes are characterized by the following theorem.
\begin{theorem}\label{teomonoqang}
The stationary, asymptotically flat solutions of the
Einstein-Maxwell equations such that in the {\em rotating tetrad} $\psi
_{0}=0$, and $\phi _{0}=0$ are given by the three-parameter family of
Kerr-Newman spacetimes. Alternatively, 
a spacetime is 
{\em monopolar with charge and angular momentum} if and only if it belongs
to the Kerr-Newman family. The following components of the null tetrad are 
given by
\begin{equation}
\xi ^{i}=\frac{\xi _{0}^{i}}{\left( r-ia\cos \left( \theta \right) \right) },
\end{equation}
\begin{equation}
X^{i}=\delta _{0}^{i}-\frac{\left( \tau _{0}\bar{\xi}_{0}^{i}+\bar{\tau}%
_{0}\xi _{0}^{i}\right) }{\left( r^{2}+a^{2}\cos ^{2}\left( \theta \right)
\right) },
\end{equation}
\begin{equation}
U=-\frac{1}{2}-\frac{\left( \tau _{0}\bar{\tau}_{0}+q\bar{q}\right) }{\left(
r^{2}+a^{2}\cos ^{2}\left( \theta \right) \right) }+\frac{Mr}{\left(
r^{2}+a^{2}\cos ^{2}\left( \theta \right) \right) },
\end{equation}
with $\xi _{0}^{0}=-\tau _{0}=i\eth _{0}\left( a\cos \left( \theta \right)
\right) $. The parameters $M$, $a$, and $q$ are the mass, angular
momentum parameter, and charge respectively.
\end{theorem}

We refer the reader to the second part of the Appendix \ref{sappendb} where
this theorem is proved.

\paragraph{Higher Momenta:}
In all the previous definitions we have required the
vanishing of the Weyl component $\psi _{0}$. The motivation for these
requirement comes from the study in linearized gravity of the multipolar
structure of compact sources. In section \ref{lili} we showed that in 
the linear
vacuum theory $\psi _{0}$ contains information concerning quadrupolar and
higher momenta of the gravitational field. In the spirit of this work, the
condition $\psi _{0}=0$ was imposed in order to exclude the possibility of
having higher momenta, and in this way, to give a model of particle in
general relativity. One can show that in the linearized Einstein-Maxwell
problem it is possible to include charge without breaking this structure (as
we did in the previous definitions). However, if one tries to introduce
higher multipolar terms in the Maxwell field one has to necessarily abandon
the condition $\psi _{0}=0$. For example, in linearized gravity, if one
includes a dipolar momentum $\mu $ in the Maxwell field (i.e., if one takes $%
\phi _{1}=\frac{\mu }{r^{3}}$) then the Bianchi identities require for $\psi
_{0}$ that $3\psi _{0}^{1}+\overline{\eth }_{0}\eth _{0}\psi _{0}^{1}=-2\eth %
_{0}\mu \eth _{0}\bar{\mu}$. In order to include a dipolar Maxwell field one
has to necessarily include higher momenta in the gravitational field.

\section{Final comments}

We have succeeded in obtaining exact solutions of Einstein equations
starting with asymptotic data with a natural physical interpretation in
linearized gravity. These data were designed to mimic the structure of well
understood solutions of linearized gravity representing compact objects,
where the presence of the Minkowski background metric provides means of
unambiguous physical interpretation. 

In the description of systems with
angular momentum the results of section \ref{rotating} show the
convenience of the notion of the {\em rotating tetrad }(a twisting null
tetrad chosen to annihilate the leading asymptotic component of $\psi _{1}$).

It is striking that the physically most relevant stationary Einstein-Maxwell
solutions representing isolated systems are recovered by means of our method.

Finally, this work suggests the possibility of generalizing the method to
non-stationary asymptotically flat spacetimes. As it is shown in reference 
\cite{Moreschi98} there is a preferred family of sections of $\cal I^{+}$, 
the {\em nice sections}, that represent the notion of instantaneous 
rest frame for
asymptotically flat spacetimes. We would like to extend our construction to
non-stationary cases by basing the null tetrads on these family of sections.
In this way we expect to be able to contribute to the construction of an 
approximation
scheme, suitable for the description of the dynamics of compact objects 
in general relativity,
by means of a model with a finite number of degrees of freedom (mass,
angular momentum, etc.).

\section*{Acknowledgments}
We acknowledge support from  Fundaci\'on YPF, 
 SeCyT-UNC, CONICET and FONCYT BID 802/OC-AR PICT: 00223.

\appendix

\section{Tetrad transformations\label{gauge}}

The freedom in choosing a null tetrad satisfying the normalization
conditions $\ell^{a}n_{a}=-m^{a}\overline{m}_{a}=1$,
where all other contractions vanish,
is given by the following transformations\cite{Prior77}:

\begin{center}
Transformation of type I
\end{center}
\[
\ell ^{a}\rightarrow \ell ^{a} ,
\]
\[
m^{a}\rightarrow m^{a}+\Gamma \ell ^{a} ,
\]
\[
n^{a}\rightarrow n^{a}+\Gamma \overline{m}^{a}+\overline{\Gamma }%
m^{a}+\Gamma \overline{\Gamma }\ell ^{a} .
\]

\begin{center}
Transformation of type II
\end{center}
\[
n^{a}\rightarrow n^{a} ,
\]
\[
m^{a}\rightarrow m^{a}+\Lambda n^{a} ,
\]
\[
\ell ^{a}\rightarrow \ell ^{a}+\Lambda \overline{m}^{a}+\overline{\Lambda }%
m^{a}+\Lambda \overline{\Lambda }n^{a}. 
\]

\begin{center}
Transformation of type III
\end{center}
\[
\ell ^{a}\rightarrow Z\ell ^{a} ,
\]
\[
n^{a}\rightarrow Z^{-1}n^{a} ,
\]
\[
m^{a}\rightarrow e^{i\theta }m^{a} ,
\]
where $\Gamma $ and $\Lambda $ are complex scalars, whereas $Z$ and $\theta $
are real. Therefore, we have six real scalar representing the degrees of
freedom of the local Lorentz rotations.

\section{General tetrad associated to a null congruence\label{sappenda}}

This appendix contains the equations relating the spin coefficients, in the
GHP formalism, with the null tetrad components associated with the coordinate
system adapted to a real null vector field $\ell ^{a}$, in the sense
explained below. The equations corresponding to the null tetrads used
throughout this paper can be simply obtained by setting certain coefficients
to zero according to the following rules. In order to obtain the equations
in the {\em standard tetrad} of section \ref{standard} set $\omega =\xi
^{0}=\rho -\bar{\rho}=\kappa =\epsilon =0$ in the equations for the
general tetrad presented in this appendix. Set $\omega =\kappa =\epsilon =0$
to obtain the relations among spin coefficients in the {\em rotating tetrad}.

The coordinate $r$ is chosen such that the vector field $\ell ^{a}$ is given
by:
\begin{equation}
\ell ^{a}=\left( \frac{\partial }{\partial \,r}\right) ^{a}.
\end{equation}

Let $\Sigma $ be a hypersurface such that the integral lines of the vector
field $\ell ^{a}$ intercept $\Sigma $ only once. Let $\left\{ \hat{t},\hat{x}%
^{2},\hat{x}^{3}\right\} $ be a coordinate system on $\Sigma$; then, in a
neighborhood of $\Sigma $ one can define coordinates $\left\{
\,t,r,x^{2},x^{3}\right\} $ such that on $\Sigma $
\begin{equation}
t=\hat{t}\quad,\quad x^{2}=\hat{x}^{2}\quad,\quad x^{3}=\hat{x}^{3},
\end{equation}
and $\left\{ \,t,x^{2},x^{3}\right\} $ are constant along the integral lines
of $\ell ^{a}$.

In this coordinate system the remaining three null vectors are given by:
\begin{equation}
m^{a}=\omega \left( \frac{\partial }{\partial r}\right) ^{a}+\xi ^{i}\left( 
\frac{\partial }{\partial x^{i}}\right) ^{a},
\end{equation}
\begin{equation}
\bar{m}^{a}=\bar{\omega}\left( \frac{\partial }{\partial r}\right) ^{a}+\bar{%
\xi}^{i}\left( \frac{\partial }{\partial x^{i}}\right) ^{a},
\end{equation}
\begin{equation}\label{vecn}
n^{a}=U\left( \frac{\partial }{\partial r}\right) ^{a}+X^{i}\left( \frac{%
\partial }{\partial x^{i}}\right) ^{a},
\end{equation}
where the index $i$ takes the values $i=0,2,3$; the vector $n^{a}$ is real
and $\bar{m}^{a}$ represents the complex conjugate of $m^{a}$.

The corresponding representation of the dual null tetrad is given by:
\begin{equation}
\ell _{a}=\frac{1}{d}\epsilon _{ijk}\xi ^{j}\bar{\xi}^{k}\left(
dx^{i}\right) _{a},
\end{equation}
\begin{equation}
m_{a}=\frac{1}{d}\epsilon _{ijk}\xi ^{j}X^{k}\left( dx^{i}\right) _{a},
\end{equation}
\begin{equation}
n_{a}=\frac{1}{d}\epsilon _{ijk}\left[ U\bar{\xi}^{j}\xi ^{k}+\omega X^{j}%
\bar{\xi}^{k}+\bar{\omega}\xi ^{j}X^{k}\right] \left( dx^{i}\right)
_{a}+\left( dr\right) _{a},
\end{equation}
where $d\equiv \epsilon _{ijk}X^{i}\xi ^{j}\bar{\xi}^{k}$, $\epsilon
_{ijk}=\epsilon _{\left[ ijk\right] }\,$ with $i,j,k=0,2,3$, and \thinspace $%
\epsilon _{023}=1$.

The torsion free condition on the covariant derivative relates the
commutator of the tetrad vectors and the spin coefficients\footnote{%
The definition of the spin coefficients in terms of the covariant derivative
of the elements of the null tetrad can be found in ref. \cite{Geroch73}.}, 
which
are:
\begin{equation}
\left[ \ell ,m\right] ^{a}=\left( \bar{\rho}+\epsilon -\bar{\epsilon}\right)
m^{a}+\sigma \bar{m}^{a}+\left( \bar{\beta}^{\prime }-\beta -\bar{\tau}%
^{\prime }\right) \ell ^{a}-\kappa n^{a},
\end{equation}

\begin{equation}
\left[ \ell ,n\right] ^{a}=\left( \epsilon ^{\prime }+\bar{\epsilon}^{\prime
}\right) \ell ^{a}+\left( \tau -\bar{\tau}^{\prime }\right) \bar{m}%
^{a}+\left( \bar{\tau}-\tau ^{\prime }\right) m^{a}-\left( \epsilon +\bar{%
\epsilon}\right) n^{a},
\end{equation}

\begin{equation}
\left[ n,m\right] ^{a}=\left( -\epsilon ^{\prime }+\bar{\epsilon}^{\prime
}+\rho ^{\prime }\right) m^{a}+\bar{\sigma}^{\prime }\bar{m}^{a}-\bar{\kappa}%
^{\prime }\ell ^{a}+\left( \beta -\bar{\beta}^{\prime }-\tau \right) n^{a},
\end{equation}

\begin{equation}
\left[ m,\bar{m}\right] ^{a}=\left( \rho -\bar{\rho}\right) n^{a}+\left( 
\bar{\rho}^{^{\prime }}-\rho ^{^{\prime }}\right) \ell ^{a}+\left( \bar{\beta%
}+\beta ^{\prime }\right) m^{a}-\left( \beta +\bar{\beta}^{\prime }\right) 
\bar{m}^{a}.
\end{equation}
Writing these expressions in terms of the coordinate basis, one obtains the
following equations:
\begin{equation}\label{53}
U_{r}=\left( \epsilon ^{^{\prime }}+\bar{\epsilon}^{^{\prime }}\right)
+\left[ \left( \tau -\bar{\tau}^{^{\prime }}\right) \bar{\omega}+\left( \bar{%
\tau}-\tau ^{^{\prime }}\right) \omega \right] -U\left( \epsilon +\bar{%
\epsilon}\right),
\end{equation}

\begin{equation}
X_{r}^{i}=\left( \tau -\bar{\tau}^{^{\prime }}\right) \bar{\xi}^{i}+\left( 
\bar{\tau}-\tau ^{^{\prime }}\right) \xi ^{i}-\left( \epsilon +\bar{\epsilon}%
\right) X^{i}  \label{eXi},
\end{equation}
\begin{equation}
\omega _{r}=\left( \bar{\rho}+\epsilon -\bar{\epsilon}\right) \omega +\sigma 
\bar{\omega}+\left( \bar{\beta}^{^{\prime }}-\beta -\bar{\tau}^{^{\prime
}}\right) -\kappa U,
\end{equation}

\begin{equation}
\xi _{r}^{i}=\left( \bar{\rho}+\epsilon -\bar{\epsilon}\right) \xi
^{i}+\sigma \bar{\xi}^{i}-\kappa X^{i}  \label{exi},
\end{equation}
\begin{multline}
U\omega _{r}+X^{k}\omega _{k}-\omega U_{r}-\xi ^{k}U_{k}= \\ 
\left( \beta -\bar{\beta}^{\prime }-\tau \right) U-\bar{\kappa}^{\prime
}-\left( \epsilon ^{\prime }-\bar{\epsilon}^{\prime }-\rho ^{\prime }\right)
\omega +\bar{\sigma }^{\prime }\bar{\omega},
\end{multline}
\begin{multline}
U\xi_{r}^{i}+X^{k}\xi _{k}^{i}-\omega X_{r}^{i}-\xi ^{k}X_{k}^{i}= \\ 
\left( \beta -\bar{\beta}^{\prime }-\tau \right) X^{i} - \left( \epsilon
^{\prime }-\bar{\epsilon}^{\prime }-\rho ^{\prime }\right) \xi ^{i} + 
\bar\sigma' \bar{\xi}^{i}
\label{co}
\end{multline}
\begin{multline}
\omega \bar{\omega}_{r}+\xi ^{p}\bar{\omega}_{p}-\bar{\omega}\omega _{r}-
\bar{\xi}^{p}\omega _{p}= \\ 
\left( \rho -\bar{\rho}\right) U+\left( \bar{\rho}^{\prime }-\rho ^{\prime
}\right) +\left( \bar{\beta}+\beta ^{\prime }\right) \omega -\left( \beta +%
\bar{\beta}^{\prime }\right) \bar{\omega}
\label{u}
\end{multline}
\begin{multline}
\omega \bar{\xi}_{r}^{i}+\xi ^{p}\bar{\xi}_{p}^{i}-\bar{\omega}\xi _{r}^{i}-
\bar{\xi}^{p}\xi _{p}^{i}= \\ 
\left( \rho -\bar{\rho}\right) X^{i}+\left( \bar{\beta}+\beta ^{\prime
}\right) \xi ^{i}-\left( \beta +\bar{\beta}^{\prime }\right) \bar{\xi}^{i}
,  \label{B=0}
\end{multline}
where the sub-indices represent the corresponding coordinate derivative. The
previous equations are the ones that allow us to integrate the components of
the null tetrad once we have found the spin coefficients that solve
Einstein equations in the GHP formalism.

Putting the spin coefficients in terms of the tetrad components, one
obtains:
\begin{equation}
\rho =\frac{1}{2d}\left[ d_{r}+\epsilon _{ijk}\left[ \left( \omega \bar{\xi}%
_{r}^{i}+\xi ^{l}\bar{\xi}_{l}^{i}-\bar{\omega}\xi _{r}^{i}-\bar{\xi}^{l}\xi
_{l}^{i}\right) -X_{r}^{i}\right] \xi ^{j}\bar{\xi}^{k}\right],
\end{equation}

\begin{equation}
\sigma =-\frac{\epsilon_{ijk}}{d}\xi _{r}^{i}\,\xi ^{j}\,X^{k},
\end{equation}

\begin{multline}
  \label{eq:tau}
  \tau = \frac{\epsilon_{ijk}}{2d}
\left[ X^i_r \left(X^j \xi^k + \omega \xi^j \bar{\xi}^k \right)
\right. +\\
\left. \xi^i_r X^j \left( \bar\omega \xi^k - \omega \bar\xi^k \right)
+ \left( \xi^l X^i_l - X^l \xi^i_l \right) \xi^j \bar\xi^k
\right]
- \frac{\omega_r}{2},
\end{multline}

\begin{equation}
\kappa =-\frac{\epsilon_{ijk}}{d}\xi _{r}^{i}\xi ^{j}\bar{\xi}^{k},
\end{equation}

\begin{multline}
\rho' = -\frac{\epsilon _{ijk}}{2d}
\left\{
 \left( \omega \bar{\xi}_{r}^{i}+\xi ^{l}\bar{\xi}_{l}^{i}-\bar{\omega}\,\xi
_{r}^{i}-\bar{\xi}^{l}\xi _{l}^{i}\right) \times 
\right.\\ 
\left( U\bar{\xi}^{j}\xi ^{k}+\omega X^{j}\bar{\xi}^{k}+\bar{\omega}\xi
^{j}X^{k}\right) +\\ 
 \left[ \left( -U\xi _{r}^{i} - X^l \xi_l^i
+\omega X_{r}^{i} + \xi ^{l}X_{l}^{i}\right)  \bar{\xi}^{j}X^{k}
+ \right. \\
\left.\left. \left( U\bar\xi _{r}^{i}+X^l \bar\xi_l^i
-\bar\omega X_{r}^{i}-\bar\xi ^{l}X_{l}^{i}\right)  \xi^{j}X^{k}
\right]
\right\} \\
-\frac{1}{2}\left( \omega \bar{\omega}_{r}+\xi ^{l}\bar{\omega}_{l}-\bar{\omega}
\omega _{r}-\bar{\xi}^{l}\omega _{l}\right)  ,
\end{multline}

\begin{equation}
\sigma' =\frac{1}{d}\epsilon _{ijk}\left( U\bar{\xi}%
_{r}^{i}+X^{l}\bar{\xi}_{l}^{i}-\bar{\omega}X_{r}^{i}-\bar{\xi}%
^{l}X_{l}^{i}\right) \bar{\xi}^{j}X^{k},
\end{equation}

\begin{multline}
  \label{eq:taup}
  \tau' = \frac{\epsilon_{ijk}}{2d}
\left[ X^i_r \left(X^j \bar\xi^k - \bar\omega \bar\xi^j \xi^k \right)
\right. +\\
\left. \xi^i_r X^j \left( \bar\omega \xi^k - \omega \bar\xi^k \right)
+ \left( \bar\xi^l X^i_l - X^l \bar\xi^i_l \right) \bar\xi^j \xi^k
\right]
- \frac{\bar\omega_r}{2},
\end{multline}

\begin{multline}
  \label{eq:kappap}
\kappa' = \bar\omega U_r +\bar\xi^k U_k  - 
\left( U \bar\omega_r + X^k \bar\omega_k  \right) +\\
 \frac{\epsilon_{ijk}}{d}
\left\{
\left[ X^i_r \bar\omega + X_l^i \bar\xi^l - \bar\xi^i_r U - \bar\xi^i_l X^l
\right] \right. \\
\left.
\left[ X^j \left( \omega \bar\xi^k - \bar\omega  \xi^k \right)
+ \bar\xi^j \xi^k U \right] \right\},
\end{multline}

\begin{multline}
\beta =\frac{1}{4d}\epsilon_{ijk} \Bigl[ 
X^i_r \left( X^j \xi^k -\xi^j \bar\xi^k \omega \right)
- X^i_l \xi^l \xi^i \bar\xi^k \Bigr. \\
+ \xi^i_r \left(3 X^j \xi^k \bar\omega - X^j \bar\xi^k \omega 
  + 2 \xi^j \bar\xi^k U\right)  
 + \xi^i_l \left( 2 X^j \bar\xi^l \xi^k + X^l \xi^j \bar\xi^k \right)
\\ \Bigl.
- 2 \bar\xi^i_r X^j \xi^k \omega - 2 \bar\xi^i_l \xi^l X^j \xi^k \Bigr]
- \frac{\omega_r}{4},
\end{multline}

\begin{multline}
\beta' =\frac{1}{4d}\epsilon_{ijk}\Bigl[ 
X^i_r \left( X^j \bar\xi^k - \bar\xi^j \xi^k \bar\omega \right)
- X^i_l \bar\xi^l \bar\xi^j \xi^k \Bigr. \\
+ 2 \xi^i_r X^j \bar\xi^k \bar\omega + 2 \xi^i_l \bar\xi^l X^j \bar\xi^k \\
\Bigl.
+ \bar\xi^i_r \left( 2 \bar\xi^j \xi^k U - X^j \xi^k \bar\omega
- X^j \bar\xi^k \omega \right)
+ \bar\xi^i_l \left( X^l \bar\xi^j \xi^k -2\xi^l X^j \bar\xi^k \right)
\Bigr] + \frac{\bar\omega_r}{4},
\end{multline}

\begin{multline}
\epsilon =\frac{\epsilon _{ijk}}{4d}
\Bigl[ -2 X^i_r \xi^j \bar\xi^k 
- \xi _{r}^{i} \left( X^j \bar{\xi}^{k} + \xi^j \bar\xi^k \bar\omega \right)
- \xi^i_l \bar\xi^l \xi^j \bar\xi^k \Bigr. \\
\Bigl. - \bar\xi _{r}^{i} \left( X^j \xi^{k} - \xi^j \bar\xi^k \omega \right)
+ \bar\xi^i_l \xi^l \xi^j \bar\xi^k \Bigr],
\end{multline}

\begin{multline}
\bar{\epsilon}' =\frac{\epsilon _{ijk}}{4d}
\Bigl\{ 
X_{r}^{i} \Bigl[ X^j \left( \bar\xi^k \omega - 3  \xi^k \bar\omega \right)
 -2\xi^j \bar\xi^k U \Bigr] 
 - X^i_l X^j \left(\xi^k \bar\xi^l + \bar\xi^k \xi^l \right) \Bigr. \\
+\xi^i_r \Bigl[ X^j \left[
 \bar\xi^k \left(U+\omega \bar\omega \right) -\xi^k \bar\omega^2 \right]
 - \xi^j \bar\xi^k U \bar\omega \Bigr] \\
+\xi^i_l \Bigl[ X^j \left(
 X^l \bar\xi^k - \bar\xi^l \xi^k \bar\omega + \bar\xi^l \bar\xi^k \omega 
 \right) - \xi^j \bar\xi^k \bar\xi^l U \Bigr] \\
+\bar\xi^i_r \Bigl[ X^j \left[
 \xi^k \left(U+\omega \bar\omega \right) -\bar\xi^k \omega^2 \right]
 + \xi^j \bar\xi^k U \omega \Bigr] \\
+\bar\xi^i_l \Bigl[ X^j \left(
 X^l \xi^k - \xi^l \bar\xi^k \omega + \xi^l \xi^k \bar\omega 
 \right) + \xi^j \bar\xi^k \xi^l U \Bigr] \\
+\frac{1}{4}\left(
2U_r + \bar\xi^l \omega_l-\xi^l \bar\omega_l +\bar\omega\omega_r 
- \omega \bar\omega_r \right),
\end{multline}

These relations, for a general null tetrad, generalize the equations of  
ref. \cite{Moreschi87} that were derived for a {\em standard tetrad}.

\section{Proof of the main theorems\label{sappendb}}

\subsection{Proof of Theorem \ref{tangb}}
In this first sub-section of the appendix we study the implications of the
definition of spacetimes which are {\em monopolar with angular momentum of
type b}. Since the definition is based on the {\em rotating tetrad}, one has
to begin with $\psi _{1}^{0}=0$. Additionally one requires the Weyl
tensor to satisfy $\psi _{0}=0,\,\,\,\psi _{2}^{0}\neq 0$. In the following
proofs we  refer to Einstein equations and Bianchi identities as
written in reference \cite{Moreschi87}.

The first two radial equations for the spin coefficients are given by Sachs
optical scalar equations 
\begin{equation}
\frac{\partial \rho }{\partial r}=\rho ^{2}+\sigma \bar{\sigma}
\quad,\quad \frac{\partial \sigma }{\partial r}=\sigma 
\left( \rho +\bar{\rho}\right) .
\end{equation}
These equations can be written in terms of the matrix $W=\left[ 
\begin{array}{cc}
\rho & \sigma \\ 
\bar{\sigma} & \bar{\rho}
\end{array}
\right] $ as 
\begin{equation}
\frac{\partial W}{\partial r}=W^{2};
\end{equation} 
whose solution is given by
\begin{equation}
W=\left[  \begin{array}{cc}
\frac{\rho _{0}-r}{R^{2}} & \frac{\sigma _{0}}{R^{2}} \\ 
\frac{\bar{\sigma}_{0}}{R^{2}} & \frac{\bar{\rho}_{0}-r}{R^{2}}
\end{array}
\right],
\end{equation}
where $R^{2}=\left( \rho _{0}-r\right) \left(
\bar{\rho}_{0}-r\right) -\sigma _{0}\bar{\sigma}_{0}$. In this case 
$\rho_{0}$ is imaginary by the conditions on the coordinate system (see
equation (\ref{ro})).

From the Bianchi identities equations ( (3.91) of ref. \cite{Moreschi87}) one
obtains
\begin{equation}
\frac{\partial \psi _{1}}{\partial r}=4\rho \psi _{1},
\end{equation}
which implies that $\psi _{1}=\psi _{1}^{0}\left( \frac{1}{R^{4}}+O\left( 
\frac{1}{R^{5}}\right) \right) $. However, in the {\em rotating tetrad} $%
\psi _{1}^{0}=0$, and thus $\psi _{1}^{{}}=0$. Therefore, we have that $%
\psi _{0}=\psi _{1}=0$. Now the Goldberg-Sachs theorem implies that $\sigma
=0$, and therefore $\rho $ becomes 
\begin{equation}\label{rwrw}
\rho =-\left( r+i\,A\right) ^{-1},
\end{equation}
where $A=-i\,\rho _{0}$ is a real function of the angular coordinates to be
determined.

From equations (3.52), (3.56) and (3.57) of ref. \cite{Moreschi87} we obtain:
\begin{equation}
\tau =\frac{\tau _{0}}{r^{2}+A^{2}},
\end{equation}
\begin{equation}
\tau' =\frac{-\bar{\tau}_{0}}{\left( r+iA\right) ^{2}},
\end{equation}
\begin{equation}
\beta =\frac{\beta _{0}}{\left( r-iA\right) },
\end{equation}
\begin{equation}
\beta
^{\prime }=\frac{\bar{\beta}_{0}}{\left( r+iA\right) }-\frac{\bar{\tau}_{0}}{%
\left( r+iA\right) ^{2}},
\end{equation}
where $\tau_0$ and $\beta_0$ are, so far, undetermined functions of the angular
coordinates. From the torsion free conditions (\ref{eXi}) and (\ref{exi}) in
the appendix \ref{sappenda} the radial dependence of the tetrad components 
$\xi^{i}$ and $X^{i}$ is given by: 
\begin{equation}
\xi ^{i}=\frac{\xi _{0}^{i}}{\left( r-iA\right) }
\quad,\quad X^{i}=X_{0}^{i}-\frac{%
\left( \tau _{0}\bar{\xi}_{0}^{i}+\bar{\tau}_{0}\xi _{0}^{i}\right) }{\left(
r^{2}+A^{2}\right) },
\end{equation}
where $\xi^i_0$ are to be determined.

The choice of asymptotic angular coordinates (see equation (\ref{x})) requires
$ X_{0}^{2}=X_{0}^{3}=0$; while the quantities $\xi _{0}^{2}$ and $\xi
_{0}^{3} $ are given by (\ref{exi0}) and (\ref{exi0p}). The value of $\xi
_{0}^{0}$ is going to be determined below.

Given a quantity $\eta $ of type $\left\{ p,q\right\} $ (see GHP formalism%
\cite{Geroch73}), one can express the action of the $\eth $-operator in 
terms of
the leading order operator $\eth _{0}$ (the $\eth $-operator on the unit
sphere), as follows:
\begin{equation}
\eth \eta =\frac{\eth _{0}\left( \eta \right) }{\left( r-iA\right) }
+q\bar{\tau}^{^{\prime }}\eta 
\quad,\quad
\overline{\eth }\eta =\frac{\overline{\eth }%
_{0}\left( \eta \right) }{\left( r+iA\right) }+p\tau ^{^{\prime }}\eta .
\end{equation}

From equation (3.55) of ref. \cite{Moreschi87} we can determine the angular
dependence of the function $\tau _{0}$ in terms of the function $A$, as
\begin{equation}
\tau _{0}=-i\eth _{0}A  \label{eethA}.
\end{equation}

From equation (3.54) of ref. \cite{Moreschi87} we calculate the value of the 
spin coefficient $\sigma'$, namely: 
\begin{equation}
\sigma' =\frac{%
\overline{\eth }_{0}\bar{\tau}_{0}}{\left( r+i\,A\right) ^{2}},
\label{sipri}
\end{equation} 
and from the
Bianchi identities (3.89) and (3.90) in ref. \cite{Moreschi87} we obtain 
\begin{equation}
\psi _{2}=\frac{\psi _{2}^{0}}{\left( r+i\,A\right) ^{3}},
\end{equation} 
together with 
\begin{equation}
\eth _{0}\psi _{2}^{0}=0.
\end{equation} 
Therefore, we can express $\psi
_{2}^{0}$ as $\psi _{2}^{0}=M+iB$ where $M$ and $B$ are two real constants.

From (3.53) of ref. \cite{Moreschi87} one obtains
\begin{equation}
\rho' =\frac{\rho _{0}'}{\left( r-iA\right) }-\frac{%
\eth _{0}\tau _{0}'}{\left( r^{2}+A^{2}\right) }+\frac{\tau
_{0}\bar{\tau}_{0}+r\left(-M-iB\right) }{\left( r+iA\right) \left(
r^{2}+A^{2}\right) }.
\end{equation}

For the case of a {\em rotating tetrad}, equation (\ref{u}) of appendix 
\ref{sappenda} is an algebraic relation
between $\rho$, $\rho'$ and $U$; from which one can obtain 
\begin{equation}
U=\rho _{0}'-\frac{\left( \eth _{0}\tau _{0}'-%
\overline{\eth }_{0}\bar{\tau}_{0}' \right) }{2iA}-\frac{%
\tau _{0}\bar{\tau}_{0}}{\left( r^{2}+A^{2}\right) }+\frac{\left(
M-\frac{r}{A}B\right) r}{\left( r^{2}+A^{2}\right) }.
\end{equation}

From equation (3.59) of ref. \cite{Moreschi87} we obtain 
$\rho_{0}'=1/2$ and 
\begin{equation}
2i\left( A+B\right) =\eth _{0}\tau _{0}'-\overline{\eth }_{0}\bar{\tau}_{0}'.
\end{equation}
This last equation implies, using (\ref{eethA}), 
\begin{equation}
\label{ee}
\eth _{0}\overline{\eth 
}_{0}A=-\left( A+B\right),
\end{equation} 
i.e. $A$ is a combination of spherical
harmonics up to $\ell =1$. Performing a suitable change of coordinates at 
$\cal I^{+}$ one can write $A$ as:
\begin{equation}\label{AA}
A=a\cos \left( \theta \right) -B;  
\end{equation}
moreover, (\ref{ee}) implies $\eth _{0}\tau _{0}=0$ 
which together with (\ref{sipri}) implies $\sigma'=0$.

From eq. (3.58) of ref. \cite{Moreschi87} one obtains 
\begin{equation}
\epsilon ^{^{\prime }}=\epsilon
_{0}^{^{\prime }}+\frac{\left( \bar{\tau}_{0}\beta _{0}-\tau _{0}\bar{\beta}%
_{0}\right) }{\left( r^{2}+A^{2}\right) }+\frac{\tau _{0}\bar{\tau}%
_{0}\left( r-iA\right) }{\left( r^{2}+A^{2}\right) ^{2}}-\frac{M+iB}{2\left(
r+iA\right) ^{2}}, 
\end{equation}
and the asymptotic analysis of equation (\ref{53}) and (\ref{co}) implies 
$\epsilon _{0}'=0$.

From the torsion free conditions (\ref{co}), in the leading order of the
asymptotic expansion, one obtains 
\begin{equation}
\xi _{0}^{0}=-\tau _{0}=i\eth _{0}A,  \label{xi00}
\end{equation}
and from (\ref{B=0}), also in the leading order, it is deduced that
\begin{equation}
\beta _{0}=-\frac{1}{\sqrt{2}}\frac{\partial }{\partial \zeta }P_{0},
\end{equation}
and 
\begin{equation}
2iAX_{0}^{0}=\eth _{0}\bar{\xi}_{0}^{0}-\overline{\eth }_{0}\xi _{0}^{0}.
\end{equation}
According to equation (\ref{x}), $X_{0}^{0}=1$; therefore, the last equation
plus  (\ref{AA}), and  (\ref{xi00}) imply $B=0$.

Straight forward calculation of $\kappa'$ gives zero. Therefore the null vector
$n$ is geodesic with zero shear ($\sigma'=0$). Then the Goldberg-Sachs theorem
implies that  $\psi_{3}=\psi_{4}=0$.
We conclude then that the spacetime is algebraically
special of type D.

Finally the value of the tetrad
components becomes: 
\begin{equation} \xi ^{i}=\frac{\xi _{0}^{i}}{\left(
r-ia\cos \left( \theta \right) \right) }, 
\end{equation}
\begin{equation}
X^{i}=\delta _{0}^{i}-\frac{\left( \tau _{0}\bar{\xi}_{0}^{i}+\bar{\tau}%
_{0}\xi _{0}^{i}\right) }{\left( r^{2}+a^{2}\cos ^{2}\left( \theta \right)
\right) },
\end{equation}
\begin{equation}
U=-\frac{1}{2}-\frac{\tau _{0}\bar{\tau}_{0}}{\left( r^{2}+a^{2}\cos
^{2}\left( \theta \right) \right) }+\frac{Mr}{\left( r^{2}+a^{2}\cos
^{2}\left( \theta \right) \right) },
\end{equation}
where $\xi _{0}^{0}=i\eth _{0}A$, and $A=a\cos \left( \theta \right) .$

This family of algebraic special solutions of type D corresponds to the Kerr
family. Notice that our tetrad is different from those appearing in the text
books by Hawking\cite{Hawking73}
and Wald\cite{Wald84} and in references by Newman \& Janis\cite{Newman65} 
and Demianski \& Newman\cite{Demianski66}.

\subsection{Proof of Theorem \ref{teomonoqang}}
The asymptotic conditions in definition \ref{defmonoq}
in section \ref{sexactmax} 
for the
electromagnetic field allows us to follow the same path in the integration
of the fields equation as showed in appendix \ref{sappenda} for the 
vacuum case, all 
up to equation (\ref{eethA}), since the Ricci components appearing in the GHP
equations up to this point all vanish due to the conditions on the Maxwell
field. With these partial results Maxwell equations\cite{Newman80} reduce to:
\begin{equation}
\frac{\partial \phi _{1}}{\partial r}-2\rho \phi _{1}=0  \label{uno},
\end{equation}
\begin{equation}
\eth \phi _{1}-2\tau \phi _{1}=0  \label{dos},
\end{equation}
\begin{equation}
\frac{\partial \phi _{2}}{\partial r}-\eth ^{\prime }\phi _{1}+2\tau ^{\prime
}\phi _{1}-\rho \phi _{2}=0  \label{tres},
\end{equation}
\begin{equation}
n^{a}\partial _{a}\phi _{1}-\eth \phi _{2}+\tau \phi _{2}-2\rho ^{\prime
}\phi _{1}=0.  \label{hard}
\end{equation}

The first three equations can be written in terms of the spin coefficients
that where found before equation (\ref{eethA}). Equations (\ref{uno}) and
(\ref{rwrw}) imply 
\begin{equation}
\phi _{1}^{{}}=\frac{q}{\left( r+i\,A\right) ^{2}};
\end{equation}
equation (\ref{dos}) tells us that $\eth _{0}q=0$, and so $q$ is a constant.
Equations  (\ref{tres}), and (\ref{rwrw}) imply 
\begin{equation}
\phi _{2}^{{}}=\frac{\phi_{2}^{0}}{r+i\,A}. 
\end{equation}
Equation (\ref{hard}) cannot be
solved at this stage since we still don't know the form of the components of
the null vector $n^{a}$. However, using the stationarity condition and the
asymptotic form of equation (\ref{hard})\footnote{%
The rest of the terms contained in this equation turn out to reduce to
identities ones we complete the integration of the spin coefficients.} we
obtain $\eth _{0}\phi _{2}^{0}=0$ which admits $\phi _{2}^{0}=0$ as the only
regular solution (since $\phi _{2}^{0}$ has spin weight $-1$ and therefore
its expansion in spherical harmonics $_sY_{lm}$ contains $l \ge 1$). Therefore,
\begin{equation} \phi _{1}^{{}}=\frac{q}{\left( r+i\,A\right)
^{2}},\,\,\,\,\,\,\phi _{2}^{{}}=0,\,\,\,\,\,\phi _{0}^{{}}=0.
\end{equation}

With this form for the electromagnetic field the rest of the equations in
the GHP formalism can be integrated in a similar manner as it was done in
the vacuum case. Additional terms appear, but the structure of the equations
remains unchanged. Moreover, the function $A$ satisfies the same equation (%
\ref{AA}) with $B=0$, i.e., $A=a\cos (\theta )$. The final result is:
\begin{equation}
\xi ^{i}=\frac{\xi _{0}^{i}}{\left( r-ia\cos \left( \theta \right) \right) },
\end{equation}
\begin{equation}
X^{i}=\delta _{0}^{i}-\frac{\left( \tau _{0}\bar{\xi}_{0}^{i}+\bar{\tau}%
_{0}\xi _{0}^{i}\right) }{\left( r^{2}+a^{2}\cos ^{2}\left( \theta \right)
\right) },
\end{equation}
\begin{equation}
U=-\frac{1}{2}-\frac{\left( \tau _{0}\bar{\tau}_{0}+q\bar{q}\right) }{\left(
r^{2}+a^{2}\cos ^{2}\left( \theta \right) \right) }+\frac{Mr}{\left(
r^{2}+a^{2}\cos ^{2}\left( \theta \right) \right) },
\end{equation}
with $\xi _{0}^{0}=i\eth _{0}\left( A\right) $. This null tetrad corresponds
to the three parameter family of Kerr-Newman solutions.


\end{document}